\documentclass[a4paper,11pt,twoside]{article}

\usepackage{amsmath}
\usepackage{epsfig}

\newcommand{\fracwithdelims}[4]{\left#1 \frac{#3}{#4} \right#2}
\newcommand{\rhob}{{\bar{\rho}}}
\newcommand{\etab}{{\bar{\eta}}}

\newcommand{\GeV}{\,\mathrm{GeV}}
\newcommand{\MeV}{\,\mathrm{MeV}}

\newcommand{\lap}[1]%
	{\boldsymbol{\lambda}_{\sect{#1}}^{\phantom{\dagger}}}

\newcommand{\tlad}[1]%
	{\tilde{\boldsymbol{\lambda}}_{\sect{#1}}^\dagger}

\newcommand{\ord}[1]{\mathcal{O}\left( #1 \right)}

\newcommand{\sect}[1]{{\mathnormal{#1}}}

\newcommand{\nohyphens}%
        {\hyphenpenalty=10000\exhyphenpenalty=10000\relax}
\newcommand{\capdef}{}
\newcommand{\mycaption}[2][\capdef]{\renewcommand{\capdef}{#2}%
        \caption[#1]{{\itshape #2}}} 
\makeatletter
\renewcommand{\fnum@table}{\textbf{\tablename~\thetable}}
\renewcommand{\fnum@figure}{\textbf{\figurename~\thefigure}}
\makeatother

\newlength{\myem}
\newcommand{\sep}[1]{#1}
\newcounter{mysubequation}[equation]
\renewcommand{\themysubequation}{\alph{mysubequation}}
\newcommand{\mytag}{\stepcounter{mysubequation}%
\tag{\theequation\protect\sep{\themysubequation}}}
\newcommand{\globallabel}[1]{\refstepcounter{equation}\label{#1}}
\makeatletter
\renewcommand{\section}{\@startsection{section}{1}{0em}{-\baselineskip}%
{\baselineskip}{\normalfont\large\bfseries}}
\renewcommand{\subsection}%
{\@startsection{subsection}{2}{0em}{-0.7\baselineskip}%
{0.7\baselineskip}{\normalfont\bfseries}}
\makeatother

\newcommand{\preprintdate}{December 1998}
\newcommand{\preprintnumber}{UCB--PTH--98/59 \\ LBNL--42574 \\
SNS--PH/98--26 \\ OUTP--98--86--P}
\newcommand{\hepnumber}{arch-ive/9812384}
\newcommand{\titletext}{Precise tests of a quark mass
texture\thanks{This work was supported in part by the U.S. Department
of Energy under Contracts DE--AC03--76SF00098, in part by the National
Science Foundation under grant PHY--95--14797 and in part by the TMR
Network under the EEC Contract No.\ ERBFMRX--CT960090.}}
\newcommand{\authortext}{ \textbf{\smallskip Riccardo Barbieri$^{\,
a}$, Lawrence J.\ Hall$^{\, b}$, \\ Andrea Romanino$^{\, c}$}%
\medskip\\
\em\normalsize 
$\mbox{}^a$ Scuola Normale Superiore and INFN, Sezione di Pisa, \\
I-56126 Pisa, Italy
\\[0.1\baselineskip] 
$\mbox{}^b$ Department of Physics and Lawrence Berkeley National
Laboratory \\ University of California, Berkeley, California 94720, USA
\\[0.1\baselineskip] 
$\mbox{}^c$ Department of Physics, Theoretical Physics, University of Oxford,
Oxford OX1 3NP, UK}
\newcommand{\abstracttext}{The relations
$|V_{ub}/V_{cb}|=(m_u/m_c)^{1/2}$ and
$|V_{td}/V_{ts}|=(m_d/m_s)^{1/2}$ between the CKM matrix elements and
the quark masses are shown to imply a remarkably precise determination
of the CKM unitarity triangle or of the Wolfenstein parameters $\rho$,
$\eta$ consistent with the data so far. We view this as a clean test
of a quark mass texture neatly arising from a hierarchical breaking of
a U(2) flavour symmetry.}

\newcommand{\makeonlyfirstpage}{%
\setlength{\topmargin}{21mm}
\setlength{\textheight}{172mm}
\renewcommand{\footnoterule}{} 
\title{%
\normalsize\hspace*{\fill}
\begin{tabular}{l}\preprintnumber\\\hepnumber\end{tabular}
\vspace{3\baselineskip}\\\huge\bfseries\titletext\smallskip}
\author{
\begin{minipage}[t]{0.8\textwidth}
\large\centering\authortext
\end{minipage}}
\date{\preprintdate}
\begin{document}
\maketitle
\thispagestyle{empty}
\begin{abstract}\large\noindent\abstracttext\end{abstract}
\end{document}}

\title{
\normalsize
\begin{tabular}[t]{l}\hepnumber\\\preprintdate\end{tabular}
\hspace*{\fill}
\begin{tabular}[t]{l}\preprintnumber\end{tabular}
\vspace{3\baselineskip}\\\LARGE\bfseries\titletext\smallskip}
\author{\begin{minipage}[t]{0.8\textwidth}
\large\centering\authortext
\end{minipage}}
\date{}

\begin{document}

\bigskip
\maketitle
\begin{abstract}\normalsize\noindent\abstracttext\end{abstract}
\normalsize\vspace{\baselineskip}
%\clearpage

\section{Introduction}

\noindent

The most striking omission in the current description of particle
physics is a theory of fermion masses and mixings. In these respects,
experiment is ahead of theory. All of the charged fermion masses are
known, with a variable precision, as are known some of the mixing
parameters in the quark sector. The knowledge of the quark mixing
parameters should improve significantly and become perhaps complete in
a near future, especially, but not only, due to $b$-physics
experiments in various facilities. Related to this is the continuous
effort to measure or put limits on all sorts of Flavour Changing
Neutral Current processes. Even in the neutrino sector, remarkable
experimental progress is being made which might ultimately lead to the
determination of the neutrino masses and mixings as well. Theory, on
the other hand, is so far mostly limited to a phenomenological
approach, based on the studies of textures.

In this paper one such texture is considered and its experimental
consequences spelled out in detail. Our motivations for doing this are
twofold. On one side the texture that we consider is clearly motivated
in an attempt to understand the flavour problem based on a
spontaneously broken U(2) flavour
symmetry~\cite{pomarol:96a,barbieri:96b,barbieri:97c,BGHR}. On the
other side, the most constraining phenomenological relations that this
texture implies are common to several different approaches to the
quark flavour problem.

The texture we consider for the mass matrices of the $U$, $D$ quarks,
$m_{U,D}$, up to irrelevant phase factors, is~\cite{BGHR}
\begin{equation}
  \label{pattern}
  \fracwithdelims{|}{|}{m^{U,D}}{m^{U,D}_{33}} = 
  \begin{pmatrix}
    \ll {\epsilon'}^2/\epsilon & \epsilon' & \ll \epsilon' \\
    \epsilon' & \epsilon & \ord{\epsilon} \\
    \ll \epsilon' & \ord{\epsilon} & 1
  \end{pmatrix}
\end{equation}
where $\epsilon'\ll\epsilon\ll1$ are parameters dependent on the quark
charge.

It was shown in Refs~\cite{hall:93a,barbieri:97c}
that~(\ref{pattern}) implies the following form of the CKM matrix
\begin{equation}
V = \left( \begin{array}{ccc}
c_{12}^D + s_{12}^D s_{12}^U e^{i\phi}   
& s_{12}^D - s_{12}^U e^{i \phi} & -s_{12}^U s \\
s_{12}^U -s_{12}^D e^{i \phi} & c_{12}^D  e^{i \phi} + s_{12}^U s_{12}^D
&  s \\ s_{12}^D s                    &  -c_{12}^Ds            
& e^{-i\phi} 
\end{array} \right)
\label{CKM}
\end{equation}
where
\globallabel{s12}
\begin{align}
s_{12}^D &= \sqrt{{m_d \over m_s}} \left( 1 - {m_d \over 2 m_s} \right)
\mytag \\
c_{12}^D &= \sqrt{1 - (s_{12}^D)^2} \mytag \\
s_{12}^U &= \sqrt{{m_u \over m_c}} \mytag
\end{align}
or, in particular, \cite{fritzsch:79a}
\globallabel{amr}
\begin{align}
\fracwithdelims{|}{|}{V_{ub}}{V_{cb}} &= \sqrt{\frac{m_u}{m_c}} \mytag
\\
\fracwithdelims{|}{|}{V_{td}}{V_{ts}} &= \sqrt{\frac{m_d}{m_s}} \mytag
\end{align}
with $m_u$ and $m_c$, as $m_d$ and $m_s$, renormalized at the same
scale. The further relation~\cite{fritzsch:78a,hall:93a}
\begin{equation}
\label{Vus}
V_{us} = s_{12}^D - s_{12}^U e^{i\phi}
\end{equation}
is consistent with observation so far but it does not lead to a
constraint on the CP-violating phase $\phi$ stronger
than~(\ref{amr}\sep{a,}\sep{b}) themselves together with the unitarity
of the CKM matrix.

To be precise, Eqs~(\ref{CKM},\ref{s12}) arise from an approximate
diagonalization of~(\ref{pattern}) and have therefore
corrections~\cite{barbieri:97c}. The biggest of such corrections is in
$V_{ub}$ or in~(\ref{amr}a). Its size, depending on the unknown
order-1 coefficients in the 23 and 32 elements of~(\ref{pattern}),
can range up to 10\%, whereas all other corrections are below
2--3\%. For this reason, a 10\% random correction to~(\ref{amr}a) is
included in the following considerations.

In Section~\ref{sec:constraint} we show that
Eqs~(\ref{amr}\sep{a,}\sep{b}) give a remarkably precise determination
of the standard CKM unitarity triangle or, in the commonly used
Wolfenstein parameterization~\cite{wolfenstein:83a}, of the parameters
$\rho,\eta$. In turn, as made explicit in
Sects~\ref{sec:comparison},~\ref{sec:potential}, this should allow a
clear comparison of these predictions with forthcoming experimental
results.

As mentioned, to the extent that
Sects~\ref{sec:constraint}--\ref{sec:potential} are only based on
Eqs~(\ref{amr}\sep{a,}\sep{b}), this analysis serves a broader scope
than the determination of the U(2)-predictions themselves. In
particular it applies to all flavour models that have~(\ref{pattern})
as their starting point.  A comparison of
Eqs~(\ref{amr}\sep{a,}\sep{b}) with present data has also been made in
ref~\cite{PRS}. This reference treats differently the information on
the light quark mass ratios, it does not attribute a theoretical error
to Eq~(\ref{amr}a) and it uses a different parametrization of the CKM
matrix~\cite{barbieri:97c,fritzsch:97a}. 

\section{Constraint in the $\rho$-$\eta$ plane}
\label{sec:constraint}

\noindent
In Eqs~(\ref{amr},\ref{Vus}), the ratios of light quark masses
$m_u/m_d$, $m_d/m_s$, $m_u/m_c$ are involved. In our analysis we use
$m_u/m_d$ and the combination~\cite{leutwyler:95a}
\begin{equation}
\label{Q}
Q = \frac{m_s/m_d}{\sqrt{1-(m_u/m_d)^2}}
\end{equation}
rather than $m_u/m_d$ and $m_d/m_s$. The reason is that chiral
perturbation theory determines $Q$ to a remarkably accuracy whereas
additional assumptions, plausible but not following from pure QCD, are
required to determine $m_u/m_d$. It turns out that $Q$ alone allows to
restrict the range of $\rho$, $\eta$ in a significant way.
Relations~(\ref{amr}\,a,b) allow to express $Q$, $m_u/m_d$, $m_c/m_s$ in
terms of $\rhob$, $\etab$, $m_c/m_s$. $\rho$, $\eta$, $A$,
$\lambda$ are the Wolfenstein parameters and $\rhob=c\rho$, $\etab=c\eta$,
where $c=\sqrt{1-\lambda^2}$. We have in fact on one hand 
\globallabel{Rud}
\begin{align}
\fracwithdelims{|}{|}{V_{ub}}{V_{cb}} &= \frac{\lambda}{c}
\sqrt{\rhob^2+\etab^2} \mytag 
\\
\fracwithdelims{|}{|}{V_{td}}{V_{ts}} &= \frac{\lambda}{c}
\sqrt{(1-\rhob)^2+\etab^2} \mytag
\end{align}
(the last one with a 2\% $\lambda^2$ correction suppressed) and, on the
other hand,
\globallabel{lmr}
\begin{gather}
\frac{m_u}{m_c} = \frac{m_u}{m_d} \frac{m_d}{m_s} \frac{m_s}{m_c} =
\frac{m_u/m_d}{Q\sqrt{1-(m_u/m_d)^2}}\frac{m_s}{m_c}, \mytag \\
\frac{m_d}{m_s} = \frac{1}{Q\sqrt{1-(m_u/m_d)^2}}. \mytag
\end{gather}
Therefore we get, from~(\ref{amr}\sep{a}) and~(\ref{amr}\sep{b})
respectively 
\globallabel{r2e}
\begin{gather}
(\rhob^2+\etab^2) = \frac{c^2}{\lambda^2
Q\sqrt{1-(m_u/m_d)^2}}\frac{m_s}{m_c}\frac{m_u}{m_d} \mytag \\
(1-\rhob)^2+\etab^2 = \frac{c^2}{\lambda^2 Q\sqrt{1-(m_u/m_d)^2}}.\mytag
\end{gather}
The random correction to~(\ref{amr}\sep{a}), at most of 10\%, can be
taken into account by attributing an effective extra error to
$m_c/m_s$ appearing in~(\ref{r2e}\sep{a}) of about 20\%, included in
the following considerations.

\begin{figure}
\begin{center}
\epsfig{file=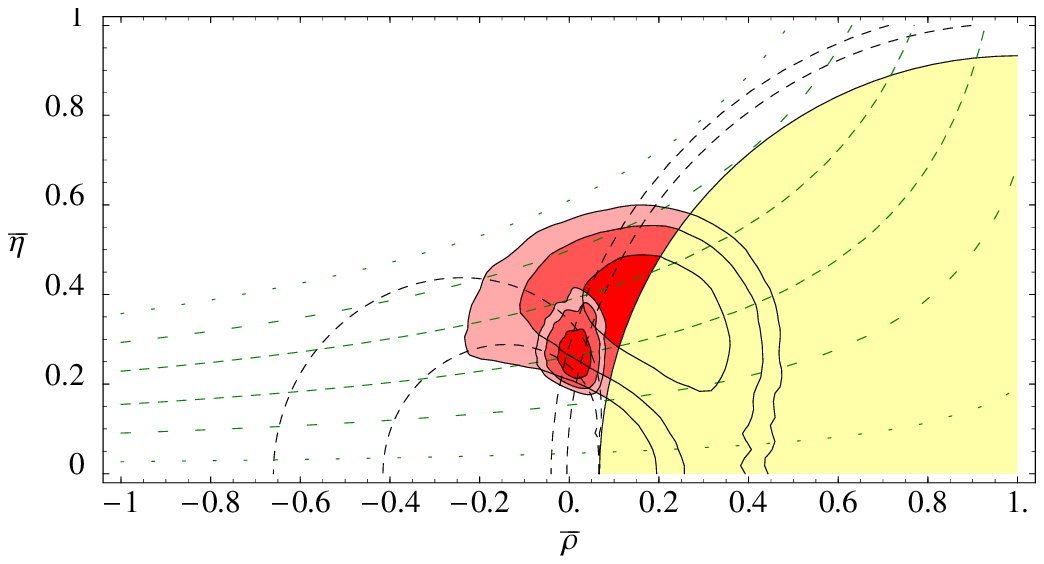,width=\textwidth}
\end{center}
\mycaption{Fit 1 (smaller regions): Predictions for $\rho$, $\eta$
using Eqs~(\ref{amr}) and the inputs in
Table~\ref{tab:inp1}. Also shown are the two individual
constraints~(\ref{r2e}\sep{b}), (\ref{r2e}\sep{c}) 
separately (see text). Fit 2 (larger regions): SM fit using
$|V_{ub}/V_{cb}|$, $\Delta m_{B_d}$, $\Delta m_{B_s}$ but not
$\epsilon_K$, whose constraint is shown independently with different
theoretical errors (see text). For both fits the contours are at 68,
95 and 99\% CL respectively.}
\label{fig:fit}
\end{figure}

From Eqs~(\ref{r2e}), using the inputs in Table~\ref{tab:inp1} we fit
the parameters $\rhob$, $\etab$. The result of the fit is shown in
Fig~(\ref{fig:fit}) (small regions).

\begin{table}
\renewcommand{\arraystretch}{1}
\[
\begin{array}{|c|c|}
\hline
Q
            & 22.7 \pm 0.8 \\  
m_u/m_d
            & 0.553 \pm 0.043 \\
m_c/m_s
            & 8.23 \pm 1.5 \\
\hline
\end{array}
\]
\mycaption{Input values for the quark mass ratios. $Q$ is defined in
the text.}
\label{tab:inp1}
\end{table}

By combining~(\ref{r2e}\sep{a}) and~(\ref{r2e}\sep{b}), it is possible
to eliminate $m_u/m_d$, whose value may not be reliably known, to
obtain
\begin{equation}
\left((1-\rhob)^2+\etab^2\right)^2
-\fracwithdelims{(}{)}{m_c}{m_s}^2 (\rhob^2+\etab^2)^2 =
\frac{c^4}{\lambda^4 Q^2}.  \mytag
\end{equation}
Fig~(\ref{fig:fit}) also shows separately the
constraints~(\ref{r2e}\sep{b}) and~(\ref{r2e}\sep{c}). The first is a
circumference in the $\rhob$-$\etab$ plane centered at
$(\rhob,\etab)=(1,0)$. The region plotted corresponds to the variation
of the right hand side of~(\ref{r2e}\sep{b}) within its $1\sigma$
error. Also shown is the region excluded by $m_u/m_d\geq 0$. The
second constraint, (\ref{r2e}\sep{c}), corresponds to the two
approximate half-circumferences around the origin meeting on the
positive $\rhob$ axes. In drawing this constraint we allow $m_c/m_s$
to vary within its 68\% error. 

Note that Eqs~(\ref{r2e}\sep{a,}\sep{b}) are invariant under
$\etab\rightarrow -\etab$, so that a complete Fig~\ref{fig:fit} would
have to include a symmetric negative $\etab$-region.

\section{Comparison with present data}
\label{sec:comparison}

\noindent
This prediction of $\rho,\eta$ can be compared with the usual
determination, in a Standard Model fit, of these same parameters. The
inputs for this last determination include the direct measurement of
$|V_{ub}/V_{cb}|$ and the indirect information from the CP-violating
parameter, $\epsilon_K$, the mixing in the $B_d$-system, $\Delta
m_{B_d}$, and in the $B_s$ system, $\Delta m_{B_s}$.  We prefer not to
include in the SM fit the parameter $\epsilon_K$ since we leave open
the possibility that some extra sources of CP-violation not included
in the SM may exist, affecting in particular CP-violation in the Kaon
system.

For ease of the reader we summarize the (standard) procedure of the SM
fit, whose result is also plotted in Fig~\ref{fig:fit} (larger
regions). The formulae used are 
\globallabel{fit2}
\begin{gather}
\fracwithdelims{|}{|}{V_{ub}}{V_{cb}} = \frac{\lambda}{c}
\sqrt{\rhob^2+\etab^2} \mytag \\
\Delta m_{B_d} = \frac{G_F^2}{6\pi} M_W^2 |V_{td}|^2 m_{B_d}
\left(f_{B_d}\sqrt{B_{B_d}}\right)^2 \eta_B\, x_t S(x_t) \mytag \\
\Delta m_{B_s} = \Delta m_{B_d} \frac{m_{B_s}}{m_{B_d}} \zeta^2
\fracwithdelims{|}{|}{V_{ts}}{V_{td}}^2, \mytag
\end{gather}
where $|V_{td}| = \lambda^3 A\, |1-\rhob-i\etab|$, $|V_{ts}|=A \lambda^2
c|1+\lambda^2(\rhob+i\etab)|$, $\zeta =(f_{B_s}\sqrt{B_{B_s}})/(f_{B_d}\sqrt{B_{B_d}})$, 
\[
S(x_t) = x_t \left( \frac{1}{4} +\frac{9}{4(1-x_t)} -\frac{3}{2(1-x_t)^2}
-\frac{3\, x_t^2\log(x_t)}{2(1-x_t)^3} \right),
\]
$x_t=m_t^2/M_W^2$ and the numerical values of the parameters are listed in
Table~\ref{tab:inp2} (in the RHS of~(\ref{fit2}c) $\Delta m_{B_d}$ is an
experimental input). The
parameters without error in Table~\ref{tab:inp2} have not been varied in the
fit. The remaining ones are quoted with their 68\% error. They are both
experimental inputs and fit variables, hence they have been integrated from the
fit by choosing them in a random way (assuming gaussian distribution). The
limit on $\Delta m_{B_s}$ has been implemented using the full set of data in
the context of the ``amplitude method'', provided by the Oscillation Working
Group~\cite{OWG}. The data correspond
to the recent limit $\Delta m_{B_s}>12.4\,\text{ps}^{-1}$ at 95\% C.L. For a
given value of $\Delta m_{B_s}$ as given by~(\ref{fit2}c), the corresponding
amplitude $a(\Delta m_{B_s})$ with its error $\sigma_a(\Delta m_{B_s})$ is
recovered from the data in~\cite{OWG}, then $a(\Delta m_{B_s})$ is
compared with 1, the value corresponding to a pure oscillation. The
multiplicative contribution to the probability density from $\Delta m_{B_s}$ is
then $\text{Exp}(-(a(\Delta m_{B_s})-1)^2/(2\sigma^2_a(\Delta m_{B_s}))$.

\begin{table}
\renewcommand{\arraystretch}{1.2}
\[
\begin{array}{||c|c||c|c||}
\hline
G_F                     & 1.16639\cdot 10^{-5}\GeV^{-2}         & 
M_W                     & 80.375                                \\
\lambda                 & 0.2196                                &
A                       & 0.819\pm 0.035                        \\
m_{B_d}                 & (5.2792\pm 0.0018)\GeV                &
m_{B_s}                 & (5.3692\pm 0.0020)\GeV                \\
f_{B_d}\sqrt{B_{B_d}}   & (0.201\pm 0.042)\GeV                  &
\zeta                   & 1.14\pm 0.08                          \\
\eta_B                  & 0.55\pm 0.01                          &
m_t                     & (166.8\pm 5.3)\GeV                    \\
\Delta m_{B_d}          & (0.471\pm 0.016)\text{ps}^{-1}        &
\epsilon_K              & (2.280\pm 0.019)\cdot 10^{-3}         \\
B_K                     & 0.87\pm 0.14                          &
f_K                     & (0.1598\pm 0.0015)\GeV                \\
m_K                     & (0.497672\pm 0.000031)\GeV            &
\Delta m_K              & (3.491\pm 0.009)\cdot 10^{-12}\MeV    \\
\eta_1                  & 1.38\pm 0.53                          &
\eta_2                  & 0.574\pm 0.004                        \\
\eta_3                  & 0.47\pm 0.04                          &
|V_{ub}/V_{cb}|         & 0.093\pm 0.016                        \\
\hline
\end{array}
\]
\mycaption{Inputs for the SM fit.}  
\label{tab:inp2}
\end{table}

Figure~\ref{fig:fit} shows consistency between the prediction
of the $\rhob,\etab$ parameters and their present determination in the
SM. The figure also shows that the constraint~(\ref{r2e}\sep{c})
alone, with the $Q$ input but without $m_u/m_d$, reduces considerably
the allowed region. In fact, a combined fit of $|V_{ub}/V_{cb}|$,
$\Delta m_{B_d}$, $\Delta m_{B_s}$ and~(\ref{r2e}\sep{c}) gives as an
output $m_u/m_d=0.61^{+0.13}_{-0.16}$.

The inclusion of $\epsilon_K$ in the SM fit would not alter the
agreement between the predicted $\rhob$, $\etab$ and present data.
To see this, the constraint given by $\epsilon_K$ is shown independently in
Fig.~\ref{fig:fit}, where the regions allowed by a SM contribution to
$\epsilon_K$ in some ranges around the central experimental
value are plotted. More precisely, the constraint is given by
\begin{equation}
\label{eps}
\frac{\epsilon_K}{C_\epsilon B_K A^2 \lambda^6} =
\etab\left(-\eta_1 S(x_c) +\eta_3 S(x_c,x_t) +A^2\lambda^4 (1-\rhob
+\lambda^2 (\rhob(1-\rhob)-\etab^2))\eta_2 S(x_t) \right),
\end{equation}
where 
\begin{gather*}
S(x_c,x_t) = x_c\left(\log\frac{x_t}{x_c}-\frac{3
x_t}{4(1-x_t)}-\frac{3x_t^2\log x_t}{4(1-x_t)^2}\right), \\
C_\epsilon=\frac{G_F^2 f_K^2 m_K M_W^2}{6\sqrt{2}\pi^2 \Delta m_K}.
\end{gather*}
The values we used for the parameters are also listed in Table~\ref{tab:inp2}
with their errors.  On the RHS in~(\ref{eps}) only the central values have been
used, that is enough for our purposes. The LHS has a 68\% error mostly coming
from the denominator. The 3 regions shown in Fig.~\ref{fig:fit} correspond to
three different ranges for the LHS all in the form $\text{LHS}\pm (68\%\text{
  error on LHS)} \pm \text{(extra error)}$, where the ``extra error'' is 0,
1/3, 2/3 of the LHS in the three regions respectively. The extra error takes
into account the possibility of contributions to $\epsilon_K$ from non-SM
physics. The figure shows that models in which relations~(\ref{amr}) are valid
do not require large corrections to the SM value.

\section{Potential improvements of the comparison with data}
\label{sec:potential}

\begin{figure}
\begin{center}
\epsfig{file=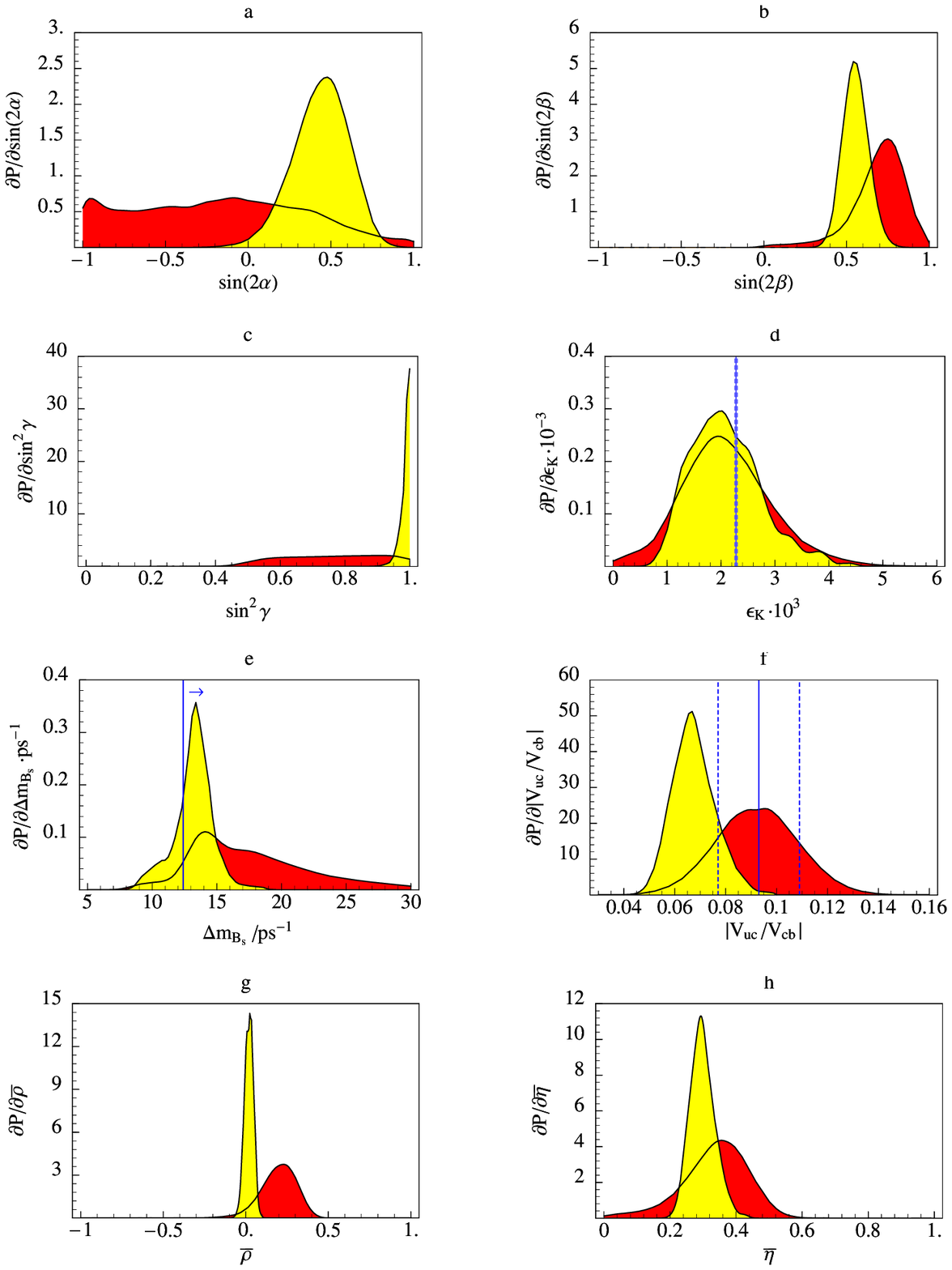,width=\textwidth}
\end{center}
\mycaption{Probability distributions (lighter area: predictions;
darker area: from the SM fit, excluding $\epsilon_K$ but taking
$\eta>0$) for different observables: a) $\sin 2\alpha$, b) $\sin
2\beta$, c) $\sin^2\gamma$, d) $\epsilon_K$, e) $\Delta m_{B_s}$, f)
$\zeta$, g) $\rho$, h) $\eta$.}
\label{fig:dis}
\end{figure}

\noindent
The prospects for an improved comparison of the predictions of the
texture~(\ref{pattern}) with further data are clear from
Fig~\ref{fig:fit}. Another way to illustrate this is in
Fig~(\ref{fig:dis}). Shown there are the probability distributions for
$\rho,\eta$ and for various physical quantities obtained by a combined
fit of $|V_{ub}/V_{cb}|$, $\Delta m_{B_d}$, $\Delta m_{B_s}$ and the
constraints~(\ref{amr}). These distributions are compared with those
from a pure SM fit with present data and $\epsilon_K$ still not
included, but rather shown as an output
(Fig~\ref{fig:dis}d)\footnote{The same distributions with the
inclusion of $\epsilon_K$ remain essentially unchanged except for
cutting away the low $\eta$-region and consequently shifting $\sin
2\alpha$ more toward zero.}.  As noted above, the sign of $\eta$ is
not fixed by Eqs~(\ref{r2e}) nor it is determined in the SM fit, since
$\epsilon_K$ is not included. As such, the probability distributions in
Figs~\ref{fig:dis}a,b,d,h, drown for $\eta>0$, must be reflected
around the origin of their orizontal axis for the $\eta<0$ case.
Based on this figure one can be optimistic on the possibility of a
stringent comparison with data to come.

\begin{figure}
\begin{center}
\epsfig{file=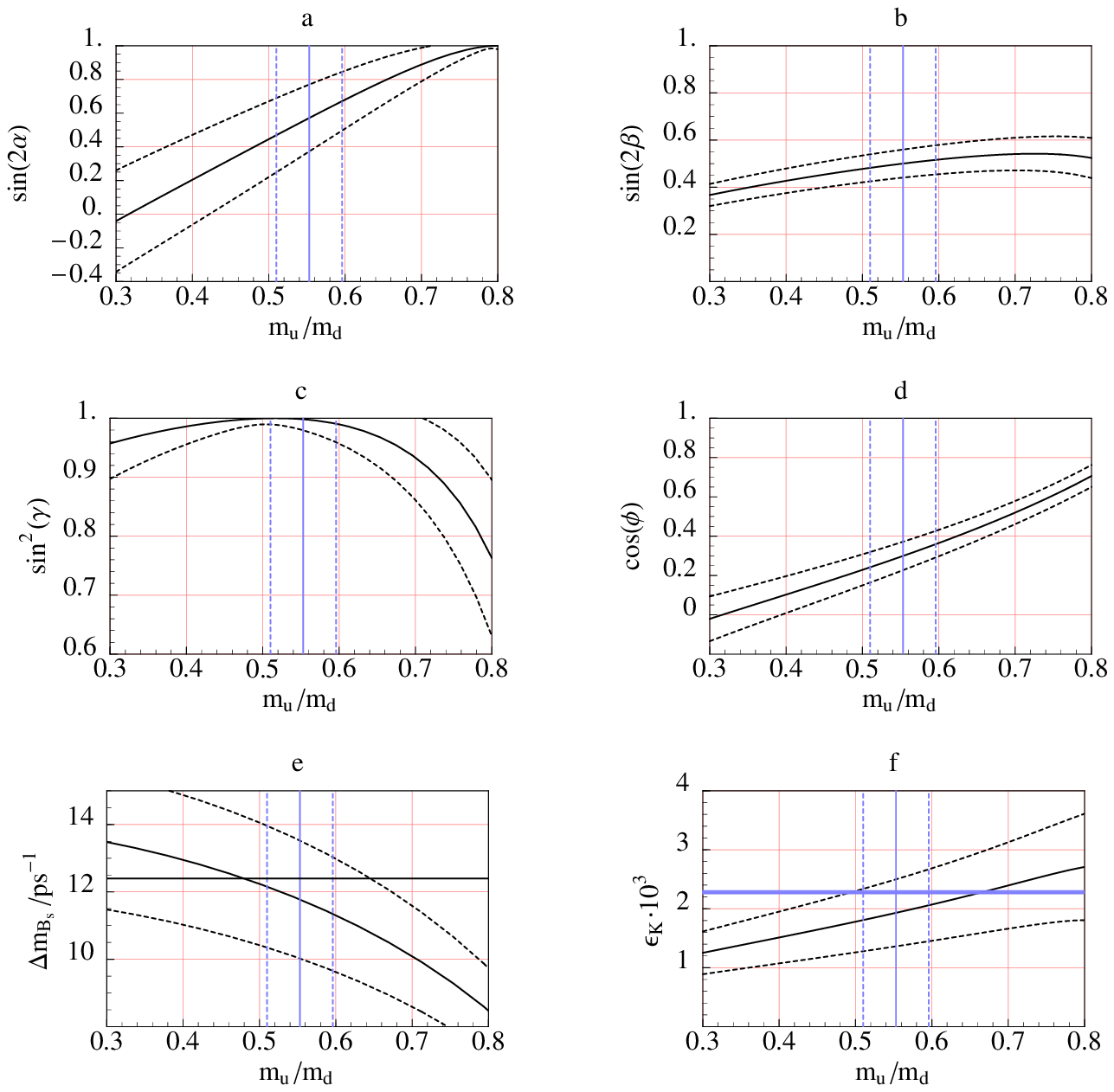,width=\textwidth}
\end{center}
\mycaption{Predictions as functions of $m_u/m_d$ for different
observables: a) $\sin 2\alpha$, b) $\sin 2\beta$, c) $\sin^2\gamma$,
d) $\cos\phi$, e) $\Delta m_{B_s}$, 
f) $\epsilon_K$.}
\label{fig:rud}
\end{figure}

It is less clear, in fact, that an improvement may come from an
independent better determination of the inputs in
Table~\ref{tab:inp1}. On the contrary, one has to keep in mind, as
repeatedly stressed, that the theoretical determination of $m_u/m_d$ is
not on the same ground as for the two other parameters. For this
reason the prediction for various physical observables is shown in
Fig~\ref{fig:rud} as function of $m_u/m_d$ and $\eta>0$. Since some of these
observables have a significant dependence on this ratio, with better
data it would be useful to leave even $m_u/m_d$ as a free
parameter. Notice in Fig~\ref{fig:dis}f the preferred value of
$|V_{ub}/V_{cb}|$ relative to the 
present determination, mostly affected by theoretical uncertainties. Notice
also in Figs~\ref{fig:dis}e and~\ref{fig:rud}e the critical lower
bound on $\Delta m_{B_s}$.

\section{Conclusions}

\noindent
The quark mass matrix texture of equation~(\ref{pattern}) provides a
simple structure of hierarchical masses and nearest neighbour
mixing. It is an immediate consequence of a hierarchical breaking of a
U(2) flavour symmetry. On diagonalizing the up and down quark matrices
with this texture, one discovers that the 12 rotations determine both
the ratio of the smaller eigenvalues and the size of 13 mixing
relative to 23 mixing: $|V_{ub} / V_{cb}| = (m_u / m_c)^{1/2}$ and
$|V_{td} / V_{ts}| = (m_d / m_s)^{1/2}$. These relations arise in any
scheme of hierarchical quark masses where the 11, 13 and 31 entries
are sufficiently small, and the 12 and 21 entries are equal up to a
phase. Using the Wolfenstein form for the CKM matrix, we have shown
that these two relations can be translated into a tight prediction in
the $\rhob$-$\etab$ plane. The predicted region is significantly
smaller than that currently allowed by data, as shown in
Figure~\ref{fig:fit}: the texture successfully accounts for the present
data and will be subject to further stringent tests by future
data. 

The most important implications for future experiments are: a
deviation from complete $B_s$ mixing must be discovered soon, $\Delta
m_{B_s} < 14.9\,\text{ps}^{-1}$ at 90\% c.l., and the predicted
probability distributions for $|\sin 2 \alpha|$ and $|\sin 2 \beta|$
are both peaked near 0.5, as shown in
Figures~\ref{fig:dis}a,b,e. These results use $m_u / m_d = 0.55 \pm
0.04$ as an input. However, even if this input is completely relaxed,
so that only one combination of the two light mass ratios is used, the
upper bound on $B_s$ mixing remains robust: $\Delta m_{B_s} < 15
\,\text{ps}^{-1}$ at 90\% c.l. In this case significant variations in
$\sin 2 \alpha$ and $\sin 2 \beta$ are possible, as shown in
Figures~\ref{fig:rud}a,b.  Note in any case that lower values of
$m_u/m_d$, that could accommodate a higher $\Delta m_{B_s}$, push
further down the expected value of $|V_{ub}/V_{cb}|$, in apparent
contradiction with the present determination.

These results are all independent of whether there is a significant
non-standard model contribution to $\epsilon_K$. However, the size of
such an exotic contribution is restricted by the form of the texture,
as shown in Figure~\ref{fig:rud}f. In the absence of such a
contribution, the combination of Figures~\ref{fig:rud}e
and~\ref{fig:rud}f predicts that $m_u/m_d$ is not far from the
commonly accepted value of $0.55 \pm 0.04$.

% \bibliographystyle{phaip}
% \bibliography{abbrev,biblio,hep,pre,tmp}

\end{document}